\documentclass[prb,aps,twocolumn,showpacs,superscriptaddress,floatfix]{revtex4}
\usepackage{epsfig}
\usepackage{amsmath}
\usepackage{amssymb}
\usepackage{amsfonts}
\usepackage{mathptmx}
\usepackage{eucal}
\usepackage{bm}
\usepackage{epstopdf}

\DeclareMathOperator{\sgn}{\mathop{\mathrm{sgn}}}
\DeclareMathOperator{\re}{\mathop{\mathrm{Re}}}
\DeclareMathOperator{\im}{\mathop{\mathrm{Im}}}

\begin{document}

\title{Theory of supercurrent transport in SIsFS Josephson junctions}
\author{S.~V.~Bakurskiy}
\affiliation{Skobeltsyn Institute of Nuclear Physics, Lomonosov Moscow State University,
Leninskie gory, Moscow 119991, Russian Federation}
\affiliation{Faculty of Physics, Lomonosov Moscow State University, Leninskie gory,
Moscow 119992, Russian Federation}
\author{N.~V.~Klenov}
\affiliation{Faculty of Physics, Lomonosov Moscow State University, Leninskie gory,
Moscow 119992, Russian Federation}
\author{I.~I.~Soloviev}
\affiliation{Skobeltsyn Institute of Nuclear Physics, Lomonosov Moscow State University,
Leninskie gory, Moscow 119991, Russian Federation}
\author{M.~Yu.~Kupriyanov}
\affiliation{Skobeltsyn Institute of Nuclear Physics, Lomonosov Moscow State University,
Leninskie gory, Moscow 119991, Russian Federation}
\author{A.~A.~Golubov}
\affiliation{Faculty of Science and Technology and MESA+ Institute for Nanotechnology,
University of Twente, 7500 AE Enschede, The Netherlands}
\date{\today }

\begin{abstract}
We present the results of theoretical study of Current-Phase Relations (CPR) $J_{S}(\varphi )$ in Josephson junctions of
SIsFS type, where 'S' is a bulk superconductor and 'IsF' is a complex weak
link consisting of a superconducting film 's', a metallic ferromagnet 'F' and an insulating barrier 'I'.
At temperatures close to critical, $T\lesssim T_{C}$, calculations are performed analytically in the frame
of the Ginsburg-Landau equations. At low temperatures numerical method is developed
to solve selfconsistently the Usadel equations in the structure.
We demonstrate that SIsFS junctions have several distinct regimes of supercurrent transport
and we examine spatial distributions of the pair potential across the structure in different regimes.
We study the crossover between these regimes which is caused by shifting the location of a weak link
from the tunnel barrier 'I' to the F-layer.
We show that strong deviations of the CPR from sinusoidal shape occur even in a vicinity of $T_{C}$, and these
deviations are strongest in the crossover regime. We demonstrate the existence of temperature-induced crossover between 0 and $\pi$ states
in the contact and show that smoothness of this transition strongly depends on the CPR shape.
\end{abstract}

\maketitle

\section{Introduction}

Josephson structures with a ferromagnetic layer became very active field of
research because of the interplay between superconducting and magnetic order
in a ferromagnet leading to variety of new effects including the realization
of a $\pi $-state with phase difference $\pi $ in the ground state of a
junction, as well as long-range Josephson coupling due generation of
odd-frequency triplet order parameter \cite{RevG,RevB,RevV}.

Further interest to Josephson junctions with magnetic barrier is due to
emerging possibilities of their practical use as elements of a
superconducting memory \cite{Oh}$^{-}$ \cite{APL}, on-chip $\pi$- phase
shifters for self-biasing various electronic quantum and classical circuits
\cite{Rogalla}$^{-}$ \cite{Ustinov}, as well as $\varphi$- batteries, the
structures having in the ground state phase difference $\varphi _{g}=\varphi
$, $(0<|\varphi |<\pi )$ between superconducting electrodes \cite%
{Buzdin,Koshelev,Pugach,Gold1,Gold2,Bakurskiy,Heim, Linder, Chan}. In
standard experimental implementations SFS Josephson contacts are
sandwich-type structures \cite{ryazanov2001}$^{-}$ \cite{Ryazanov2006a}. The
characteristic voltage $V_{C}=J_{C}R_{N}$ ($J_{C}$ is critical current of
the junction, $R_{N}$ is resistance in the normal state) of these SFS
devices is typically quite low, which limits their practical applications.
In SIFS structures \cite{Kontos}$^{-}$ \cite{Weides3} containing an
additional tunnel barrier I, the $J_{C}R_{N}$ product in a $0$-state is
increased \cite{Ryazanov3}, however in a $\pi $-state $V_{C}$ is still too
small \cite{Vasenko,Vasenko1} due to strong suppression of the
superconducting correlations in the ferromagnetic layer.

Recently, new SIsFS type of magnetic Josepshon junction was realized
experimentally \cite{Ryazanov3,Larkin,Vernik,APL}. This structure represents
a connection of an SIs tunnel junction and an sFS contact in series. Properties
of SIsFS structures are controlled by the thickness of s layer $d_{s}$ and
by relation between critical currents $J_{CSIs}$ and $J_{CsFS}$ of their SIs
and sFS parts, respectively. If the thickness of s-layer $d_{s}$
is much larger than its coherence length $\xi _{S}$ and $J_{CSIs}\ll
J_{CsFS} $, then characteristic voltage of an SIsFS device is determined by
its SIs part and may reach its maximum corresponding to a standard SIS
junction. At the same time, the phase difference $\varphi $ in a ground state
of an SIsFS junction is controlled by its sFS part.
As a result, both $0 $- and $\pi $-states can be achieved depending on a thickness of the F layer.
This opens the possibility to realize controllable $\pi $ junctions
having large $J_{C}R_{N}$ product. At the same time, being placed in
external magnetic field $H_{ext}$ SIsFS structure behaves as a single
junction, since $d_{s}$ is typically too thin to screen $H_{ext}$. This
provides the possibility to switch $J_{C}$ by an external field.

However, theoretical analysis of SIsFS junctions was not performed up to now. The
purpose of this paper is to develop a microscopic theory providing the
dependence of the characteristic voltage on temperature $T$, exchange energy $H$ in a ferromagnet,
transport properties of FS and sF interfaces and thicknesses of s and F layers.
Special attention will be given to
determining the current-phase relation (CPR) between the supercurrent $J_{S}$ and the phase
difference $\varphi $ across the structure.
\begin{figure}[tbh]
\begin{center}
\includegraphics[width=8.5cm]{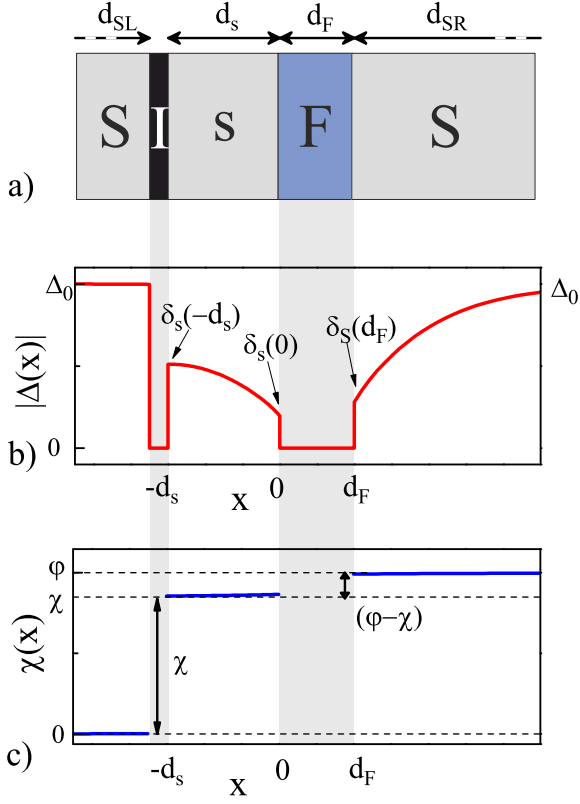}
\end{center}
\caption{ a) Schematic design of SIsFS Josephson junction. b), c) Typical
distribution of amplitude $|\Delta (x)|$ and phase difference $\protect\chi %
(x)$ of pair potential along the structure. }
\label{design}
\end{figure}

\section{Model of SIsFS Josephson device \label{Model}}

We consider multilayered structure presented in Fig.\ref{design}a. It consists
of two superconducting electrodes separated by complex interlayer including
tunnel barrier I, intermediate superconducting s and ferromagnetic F films.
We assume that the conditions of a dirty limit are fulfilled for all
materials in the structure. In order to simplify the problem, we also assume that all
superconducting films are identical and can be described by
a single critical temperature $T_{C}$ and coherence length $\xi _{S}.$
Transport properties of both sF and FS interfaces are also assumed identical
and are characterized by the interface parameters
\begin{equation}
\gamma =\frac{\rho _{S}\xi _{S}}{\rho _{F}\xi _{F}},\quad \gamma _{B}=\frac{%
R_{BF}\mathcal{A}_{B}}{\rho _{F}\xi _{F}}.  \label{gammas}
\end{equation}%
Here $R_{BF}$ and $\mathcal{A}_{B}$ are the resistance and area of the sF
and FS interfaces $\xi _{S}$ and $\xi _{F}$ are the decay lengths of S and F
materials while $\rho _{S}$ and $\rho _{F}$ are their resistivities.

Under the above conditions the problem of calculation of the critical
current in the SIsFS structure reduces to solution of the set of the Usadel
equations\cite{Usadel}. For the S layers these equations have the form \cite{RevG,
RevB, RevV}
\begin{equation}
\frac{\xi _{S}^{2}}{\Omega G_{m}}\frac{d}{dx}\left( G_{m}^{2}\frac{d}{dx}%
\Phi _{m}\right) -\Phi _{m}=-\Delta _{m},~G_{m}=\frac{\Omega }{\sqrt{\Omega
^{2}+\Phi _{m}\Phi _{m}^{\ast }}},  \label{fiS}
\end{equation}%
\begin{equation}
\Delta _{m}\ln \frac{T}{T_{C}}+\frac{T}{T_{C}}\sum_{\omega =-\infty
}^{\infty }\left( \frac{\Delta _{m}}{\left\vert \Omega \right\vert }-\frac{%
\Phi _{m}G_{m}}{\Omega }\right) =0,  \label{delta}
\end{equation}%
where $m=S$ for $x\leq -d_{s}~$and$~x\geq d_{F};$ $m=s$ in the interval $%
-d_{s}\leq x\leq 0.$ In the F film $(0\leq x\leq d_{F})$ they are
\begin{equation}
\xi _{F}^{2}\frac{d}{dx}\left( G_{F}^{2}\frac{d}{dx}\Phi _{F}\right) -%
\widetilde{\Omega }\Phi _{F}G_{F}=0.  \label{FiF}
\end{equation}%
Here $\Omega =T(2n+1)/T_{C}$ are Matsubara frequencies normalized to $\pi
T_{C}$, $\widetilde{\Omega }=\Omega +iH/\pi T_{C},$ $G_{F}=\widetilde{\Omega
}/(\widetilde{\Omega }^{2}+\Phi _{F,\omega }\Phi _{F,-\omega }^{\ast
})^{1/2},$ $H$ is exchange energy, $\xi _{S,F}^{2}=(D_{S,F}/2\pi T_{C})$ and
$D_{S,F},$ are diffusion coefficients in S and F metals, respectively. \
Pair potential $\Delta _{m}$ and the Usadel functions $\Phi _{m}$ and $\Phi
_{F}$ in (\ref{fiS}) - (\ref{FiF}) are also normalized to $\pi T_{C}.$ To
write equations (\ref{fiS}) - (\ref{FiF}), we have chosen the $x$ axis in
the directions perpendicular to the SI, FS and sF interfaces and put the
origin at sF interface. Equations (\ref{fiS}) - (\ref{FiF}) must be
supplemented by the boundary conditions \cite{KL}. At $x=-d_{s}$ they can be
written as
\begin{gather}
G_{S}^{2}\frac{d}{dx}\Phi _{S}=G_{s}^{2}\frac{d}{dx}\Phi _{s},  \label{BCIF}
\\
\gamma _{BI}\xi _{S}G_{s}\frac{d}{dx}\Phi _{s}=-G_{S}\left( \Phi _{S}-\Phi
_{s}\right) ,  \notag
\end{gather}%
where $\gamma _{BI}=R_{BI}\mathcal{A}_{B}/\rho _{S}\xi _{S}$, $R_{BI}$ and $%
\mathcal{A}_{B}$ are resistance and area of SI interface. At $x=0$ the
boundary conditions are%
\begin{gather}
\frac{\xi _{S}}{\Omega }G_{s}^{2}\frac{d}{dx}\Phi _{s}=\gamma \frac{\xi _{F}%
}{\widetilde{\Omega }}G_{F}^{2}\frac{d}{dx}\Phi _{F},  \label{BCsF} \\
\gamma _{B}\xi _{F}G_{F}\frac{d}{dx}\Phi _{F}=-G_{s}\left( \frac{\widetilde{%
\Omega }}{\Omega }\Phi _{s}-\Phi _{F}\right)  \notag
\end{gather}%
and at $x=d_{F}$ they have the form
\begin{gather}
\frac{\xi _{S}}{\Omega }G_{S}^{2}\frac{d}{dx}\Phi _{S}=\gamma \frac{\xi _{F}%
}{\widetilde{\Omega }}G_{S}^{2}\frac{d}{dx}\Phi _{F},  \label{BCSF} \\
\gamma _{B}\xi _{F}G_{F}\frac{d}{dx}\Phi _{F}=G_{S}\left( \frac{\widetilde{%
\Omega }}{\Omega }\Phi _{S}-\Phi _{F}\right) ,  \notag
\end{gather}%
Far from the interfaces the solution should cross over to a uniform
current-carrying superconducting state\cite{Ivanov}$^{-}$\cite{KuperD}%
\begin{equation}
\Phi _{S}(\mp \infty )=\Phi _{\infty }\exp \left\{ i(\chi (\mp \infty
)-ux/\xi _{S})\right\} ,~  \label{BCInf1}
\end{equation}%
\begin{equation}
\Delta _{S}(\mp \infty )=\Delta _{0}\exp \left\{ i(\chi (\mp \infty )-ux/\xi
_{S})\right\} ,  \label{BCINFD}
\end{equation}%
\begin{equation}
\Phi _{\infty }=\frac{\Delta _{0}}{1+u^{2}/\sqrt{\Omega ^{2}+|\Phi _{S}|^{2}}%
},  \label{BCinf2}
\end{equation}%
resulting in order parameter phase difference across the structure equal to%
\begin{equation}
\varphi =\varphi (\infty )-2ux/\xi _{S},~\varphi (\infty )=\chi (\infty
)-\chi (-\infty ).  \label{PhDifInf}
\end{equation}%
Here $\varphi (\infty )$ is the asymptotic phase difference
across the junction, $\Delta _{0}$ is modulus of order parameters far
from the boundaries of the structure at a given
temperature, $u=2mv_{s}\xi _{S},$ $m$ is the electron mass and $v_{s}$ is
the superfluid velocity. Note that since the boundary conditions (\ref{BCIF}%
) - (\ref{BCsF}) include the Matsubara frequency $\Omega $, the phases of $%
\Phi _{S}$ functions depend on $\Omega $ and are different from the
phase of the pair potential $\Delta _{S}$ at the FS interfaces $\chi (d_{F})$
and $\chi (0).$ Therefore it is the value $\varphi (\infty )$ rather than $%
\varphi =\chi (d_{F})-\chi (0),$ that can be measured experimentally by
using a scheme compensating the linear in $x$ part in Eq.~(\ref{PhDifInf}).

The boundary problem (\ref{fiS})-(\ref{PhDifInf}) can be solved numerically making use of (\ref{BCInf1}), (\ref{BCinf2}).
Accuracy of calculations can be monitored by equality of currents $J_{S}$
\begin{equation}
\frac{2eJ_{S}(\varphi )}{\pi T\mathcal{A}_{B}}=\sum\limits_{\omega =-\infty
}^{\infty }\frac{iG_{m,\omega }^{2}}{\rho _{m}\widetilde{\Omega }^{2}}\left[
\Phi _{m,\omega }\frac{\partial \Phi _{m,-\omega }^{\ast }}{\partial x}-\Phi
_{m,-\omega }^{\ast }\frac{\partial \Phi _{m,\omega }}{\partial x}\right] ,
\label{current}
\end{equation}%
calculated at the SI and FS interfaces and in the electrodes.

In the further analysis carried out below we limit ourselves to the
consideration of the most relevant case of low-transparent tunnel barrier at
SI interface%
\begin{equation}
\gamma _{BI}\gg 1.  \label{LargeGammaBI}
\end{equation}%
In this approximation, the junction resistance $R_{N}$ is fully determined
by the barrier resistance $R_{BI}$. Furthermore the current flowing through
the electrodes can lead to the suppression of superconductivity only in the
vicinity of sF and FS interfaces. That means, up to terms of the order of $%
\gamma _{BI}^{-1}$ we can neglect the effects of suppression
of superconductivity in the region $x\leq -d_{s}~$ and write the solution in
the form
\begin{equation}
\Phi _{S}(x)=\Delta _{S}(x)=\Delta _{0}.  \label{Sol_Left}
\end{equation}%
Here without any lost of generality we put $\chi (-\infty )=\chi
(-d_{s}-0)=0 $ (see Fig. \ref{design}c).

Substitution of (\ref{Sol_Left}) into boundary conditions (\ref{BCIF}) gives%
\begin{equation}
\gamma _{BI}\xi _{S}G_{s}\frac{d}{dx}\Phi _{s}=-\frac{\Omega }{\sqrt{\Omega
^{2}+\Delta _{0}^{2}}}\left( \Delta _{0} -\Phi _{s}\right) .  \label{BCIFmod}
\end{equation}%
Further simplifications are possible in a several limiting cases.

\section{The high temperature limit $T\approx T_{C}$\label{SecHighTC}}

In a vicinity of critical temperature the Usadel equations in the F layer
can be linearized. Writing down their solution in the analytical form and
using the boundary conditions (\ref{BCsF}), (\ref{BCSF}) on sF and FS
interfaces we can reduce the problem to the solution of Ginzburg-Landau (GL)
equations in the s and S layers. We limit our analysis by considering the
most interesting case when the following condition is fulfilled:
\begin{equation}
\Gamma _{BI}=\frac{\gamma _{BI}\xi _{S}}{\xi _{S}(T)}\gg 1,  \label{GammaBB1}
\end{equation}%
and when there is strong suppression of superconductivity in the vicinity
of the sF and FS interfaces. The latter takes place if the parameter $\Gamma $
\begin{equation}
\Gamma =\frac{\gamma \xi _{S}(T)}{\xi _{S}},~\xi _{S}(T)=\frac{\pi \xi _{S}}{%
2\sqrt{1-T/T_{C}}}  \label{GammasB1}
\end{equation}%
satisfies the conditions
\begin{equation}
\Gamma p\gg 1,~\Gamma q\gg 1.  \label{condSup}
\end{equation}%
Here
\begin{eqnarray}
p^{-1} &=&\frac{8}{\pi ^{2}}\re\sum_{\omega =0}^{\infty }\frac{1}{\Omega ^{2}%
\sqrt{\widetilde{\Omega }}\coth \frac{d_{F}\sqrt{\widetilde{\Omega }}}{2\xi
_{F}}},  \label{Sums11} \\
q^{-1} &=&\frac{8}{\pi ^{2}}\re\sum_{\omega =0}^{\infty }\frac{1}{\Omega ^{2}%
\sqrt{\widetilde{\Omega }}\tanh \frac{d_{F}\sqrt{\widetilde{\Omega }}}{2\xi
_{F}}}.  \label{Sums12}
\end{eqnarray}

Note that in the limit $h=H/\pi T_{C}\gg 1$ and
$d_{F}\gg \sqrt{2/h}\xi _{F}$ the sums in (\ref{Sums11}), (\ref{Sums12}) can
be evaluated analytically resulting in
\begin{equation}
\beta =\frac{p-q}{p+q}=\sqrt{8}\sin \left( \frac{d_{F}}{\xi _{F}}\sqrt{\frac{%
h}{2}}+\frac{3\pi }{4}\right) \exp \left( -\frac{d_{F}}{\xi _{F}}\sqrt{\frac{%
h}{2}}\right) ,  \label{pq1}
\end{equation}%
\begin{equation}
p+q=2\sqrt{2h}\left( T/T_{C}\right) ^{2},~\ \ \ pq=2h\left( T/T_{C}\right)
^{4}.\ \   \label{pq2}
\end{equation}

In general, the phases of the order parameters in s and S films are
functions of the coordinate $x$. In the considered approximation the terms
that take into account the coordinate dependence of the phases, are
proportional to small parameters $(\Gamma q)^{-1}$ and $(\Gamma p)^{-1}$ and
therefore provide small corrections to the current. For this reason, in the first approximation we can
assume that the phases in superconducting electrodes are constants
independent of x. In the further analysis we denote the phases at the s-film
by $\chi $ and at the right S-electrode by $\varphi $ (see Fig.\ref{design}%
c).

The details of calculations are summarized in the Appendix \ref{Appendix}. These
calculations show that the considered SIsFS junction has two modes of
operation depending on relation between s layer thickness $d_{s}$ and the
critical thickness $d_{{sc}}=(\pi /2)\xi _{S}(T).$ For $d_{s}$ larger than $%
d_{sc}$, the s-film keeps its intrinsic superconducting properties (\emph{%
mode (1)}), while for $d_{s}\leq d_{sc}$ superconductivity in the s-film exists only
due to proximity effect with the bulk S electrodes (\emph{mode (2)}).

\subsection{Mode (1): SIs $+$ sFS junction $d_{s}\geq d_{sc}$}

We begin our analysis with the regime when the intermediate s-layer is
intrinsically superconducting. In this case it follows from the solution of
GL equations that supercurrent flowing across SIs, sF and FS interfaces ($%
J(-d_{s}),$ $J(0)$ and $J(d_{F}),$ respectively) can be represented in the
form (see Appendix \ref{Appendix})
\begin{equation}
\frac{J_{S}(-d_{s})}{J_{G}}=\frac{\delta _{s}(-d_{s})}{\Gamma _{BI}\Delta
_{0}}\sin \left( \chi \right) ,~J_{G}=\frac{\pi \Delta _{0}^{2}\mathcal{A}%
_{B}}{4e\rho _{S}T_{C}\xi _{S}(T)},  \label{JatIS1}
\end{equation}%
\begin{equation}
\frac{J_{S}(0)}{J_{G}}=\frac{J_{S}(d_{F})}{J_{G}}=\frac{\Gamma (p-q)}{%
2\Delta _{0}^{2}}\delta _{s}(0)\delta _{S}(d_{F})\sin \left( \varphi -\chi
\right) ,  \label{JatSF1}
\end{equation}%
where $\Delta _{0}=\sqrt{8\pi ^{2}T_{C}(T_{C}-T)/7\zeta (3)}$ is bulk value
of order parameter in S electrodes, $\mathcal{A}_{B}$ is cross sectional
area of the structure, $\zeta (z)$ is Riemann zeta function. Here
\begin{equation}
\delta _{s}(0)=\frac{2b\left( p-q\right) \cos \left( \varphi -\chi \right)
-2a\left( p+q\right) }{\Gamma \left[ \left( p+q\right) ^{2}-\left(
p-q\right) ^{2}\cos ^{2}\left( \varphi -\chi \right) \right] },
\label{delta1(0)}
\end{equation}%
\begin{equation}
\delta _{S}(d_{F})=\frac{2b\left( p+q\right) -2a\left( p-q\right) \cos
\left( \varphi -\chi \right) }{\Gamma \left( \left( p+q\right) ^{2}-\left(
p-q\right) ^{2}\cos ^{2}\left( \varphi -\chi \right) \right) },
\label{delta1(df)}
\end{equation}%
are the order parameters at sF and FS interfaces, respectively (see Fig. \ref%
{design}b) and
\begin{equation}
a=-\delta _{s}(-d_{s})\sqrt{1-\frac{\delta _{s}^{2}(-d_{s})}{2\Delta _{0}^{2}%
}},~b=\frac{\Delta _{0}}{\sqrt{2}},  \label{derivaties1}
\end{equation}%
where $\delta _{s}(-d_{s})$ is the solution of transcendental equation
\begin{equation}
K\left( \frac{\delta _{s}(-d_{s})}{\Delta _{0}\eta }\right) =\frac{d_{s}\eta
}{\sqrt{2}\xi _{s}(T)},~\eta =\sqrt{2-\frac{\delta _{s}^{2}(-d_{s})}{\Delta
_{0}^{2}}}.  \label{EqDsmall1}
\end{equation}%
Here, $K(z),$ is the complete elliptic integral of the first kind.
Substitution of $\delta _{s}(-d_{s})=0$ into Eq. (\ref{EqDsmall1}) leads to
the expression for critical s layer thickness, $d_{{sc}}=(\pi /2)\xi _{S}(T)$, which was used above.

For the calculation of the CPR we
need to exclude phase $\chi $ of the intermediate s layer from the
expressions for the currents (\ref{JatIS1}), (\ref{JatSF1}). The value of
this phase is determined from the condition that the currents flowing across
Is and sF interfaces should be equal to each other.

For large thickness of the middle s-electrode ($d_{s}\gg d_{sc}$) the
magnitude of order parameter $\delta _{s}(-d_{s})$ is close to that of a bulk
material $\Delta _{0}$ and we may put $a=-b$ in Eqs.(\ref{delta1(0)}) and (%
\ref{delta1(df)})
\begin{equation}
\delta _{S}(d_{F})=\delta _{s}(0)=\frac{\sqrt{2}\Delta _{0}}{\Gamma \left(
\left( p+q\right) -\left( p-q\right) \cos \left( \varphi -\chi \right)
\right) },  \label{Aeqb1}
\end{equation}%
resulting in
\begin{equation}
J_{S}(0)=J_{S}(d_{F})=\frac{J_{G}\beta \sin \left( \varphi -\chi \right) }{%
\Gamma \left( 1-\beta \cos \left( \varphi -\chi \right) \right) }
\label{Jsymm1}
\end{equation}%
together with the equation to determine $\chi $
\begin{equation}
\frac{\Gamma }{\Gamma _{BI}}\sin \left( \chi \right) =\frac{\beta \sin
\left( \varphi -\chi \right) }{1-\beta \cos \left( \varphi -\chi \right) }%
,~\beta =\frac{p-q}{p+q}.  \label{EqForKhi1}
\end{equation}

From (\ref{Aeqb1}), (\ref{Jsymm1}) and (\ref{EqForKhi1}) it follows that in
this mode SIsFS structure can be considered as a pair of SIs and sFS
junctions connected in series. Therefore, the properties of the structure
are almost independent on thickness $d_{s}$ and are determined by a
junction with smallest critical current.

Indeed, we can conclude from (\ref{EqForKhi1}) that the phase $\chi $ of s
layer order parameter depends on the ratio of the critical current, $%
I_{CSIs}\varpropto \Gamma _{BI}^{-1},\ $of its SIs part to that, $%
I_{CsFS}\varpropto |\beta |\Gamma ^{-1},\ $of the sFS junction. The coefficient $%
\beta $ in (\ref{EqForKhi1}) is a function of F layer thickness, which becomes
close to unity in the limit of small $d_{F}$ and exhibits damped
oscillations with $d_{F}$ increase (see analytical expression for $\beta $ (%
\ref{pq1})). That means that there is a range of thicknesses, $d_{Fn,\text{ }}$%
determined by the equation $\beta =0$, at which $J_{S}\equiv 0$ and there is a
transition from $0$ to $\pi $ state in sFS part of SIsFS junction. In other
words, crossing the value $d_{Fn}$ with an increase of $d_{F}$ provides a $\pi $
shift of $\chi $ relative to the phase of the S electrode.
\begin{figure}[h]
\center{\includegraphics[width=1\linewidth]{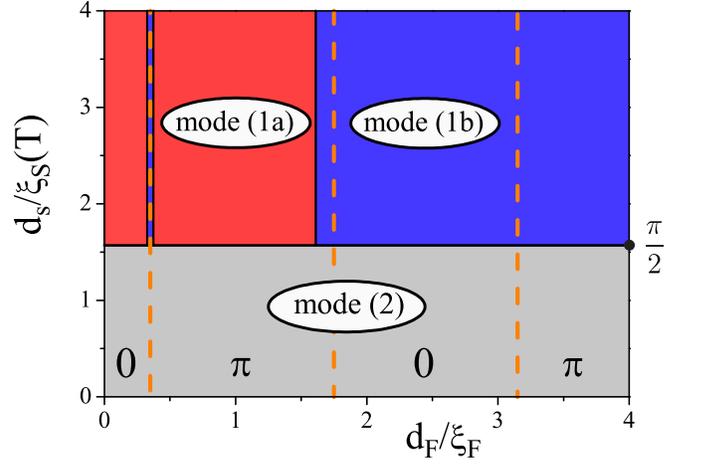}}
\caption{ The phase diagram of the operation modes of the SIsFS structure in
the $(d_{s},d_{F})$ plane. The bottom area corresponds to the \emph{mode (2)}
with fully suppressed superconductivity in the s-layer. The top part of the
diagram, separated from the bottom one by the solid horizontal line, corresponds to the s-layer in superconductive state.
It provides the value of s layer
critical thickness, $d_{sc}.$ The upper-left part indicates the \emph{mode (1a)%
} with the weak place located at SIs tunnel barrier. The upper-right area as
well as thin valley around first $0-\protect\pi $ transition correspond to
the \emph{mode (1b)} with the weak place located at sFS junction. Solid
vertical lines provide loci of the boarders between the \emph{modes (1a)} and
\emph{(1b).} Vertical dashed lines show positions of $0-\protect\pi $
transitions. The calculations have been done for $H=10\protect\pi T_{C}$, $%
\Gamma _{BI}=200$ and $\Gamma =5$ at $T=0.9T_{C}$. }
\label{Phase}
\end{figure}

In Fig.\ref{Phase} we clarify the classification of operation modes and
demonstrate the phase diagram in the $(d_{s},d_{F})$ plane, which follows
from our analytical results (\ref{pq1})-(\ref{EqDsmall1}). The calculations
have been done at $T=0.9T_{C}$ for $h=H/\pi T_{C}=10$, $\Gamma _{BI}=200$
and $\Gamma =5.$ The structures with s-layer smaller than critical thickness
\ $d_{sc}=\pi \xi _{S}(T)/2$ correspond to the \emph{mode (2)} with fully
suppressed superconductivity in the s layer. Conversely, the top part of
diagram corresponds to s-layer in the superconductive state (\emph{mode (1)}%
). This area is divided into two parts depending on whether the weak place
located at the tunnel barrier I (\emph{mode(1a)}) or at the ferromagnetic
F-layer (\emph{mode(1b)}). The separating black solid vertical lines in the
upper part in Fig.\ref{Phase} represent the locus of points where the
critical currents of SIs and sFS parts of SIsFS junction are equal. The
dashed lines give the locations of the points of $0$ to $\pi $ transitions, $%
d_{Fn}=\pi (n-3/4)\xi _{F}\sqrt{2/h},~n=1,2,3...,$ at which $J_{s}=0.$ In a
vicinity of these points there are the valleys of \emph{mode (1b)} with the
width, $\Delta d_{Fn}\approx \xi _{F}\Gamma \Gamma _{BI}^{-1}h^{-1/2}\exp
\{\pi (n-3/4)\},$ embedded into the areas occupied by \emph{mode (1a)}. For
the set of parameters used for calculation of the phase diagram presented in
Fig.\ref{Phase}, there is only one valley with the width $\Delta
d_{F1}\approx \xi _{F}\Gamma \Gamma _{BI}^{-1}h^{-1/2}\exp \{\pi /4\}$
located around the point $d_{F1}=(\pi /4)\xi _{F}\sqrt{2/h}$ of the first $0$
to $\pi $ transition.
\begin{figure}[h]
\center{\includegraphics[width=1\linewidth]{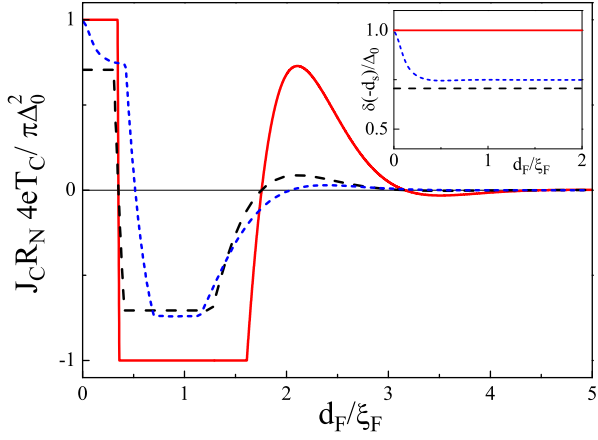}}
\caption{Critical current $J_{C}$ of the SIsFS structure versus F-layer
thickness $d_{F}$ calculated at $T=0.9T_{C},$ $H=10\protect\pi T_{C}$, $%
\Gamma _{BI}=200$ and $\Gamma =5$ for s layer thickness $d_{s}=2\protect\xi %
_{s}(T)$ slightly above the critical one $d_{sc}$. Inset shows
dependence of pair potential $\protect\delta _{s}(-d_{s})$ at the Is
interface of the s-layer versus F-layer thickness $d_{F}$. Solid lines have
been calculated for $d_{s}\gg d_{sc}$ from Eqs. (\protect\ref{SISIotFI3})-(%
\protect\ref{SInFSotFI}). The dashed line is the result of calculations
using analytical expressions (\protect\ref{JatIS1})-(\protect\ref{EqDsmall1}%
) for the thickness of s-layer $d_{s}=2\protect\xi _{s}(T).$ Short-dashed
line is the result of numerical calculations in the frame of Usadel
equations (\protect\ref{fiS})-(\protect\ref{PhDifInf}). }
\label{Ic_ds}
\end{figure}

\subsubsection{Mode (1a): Switchable $0-\protect\pi $ \ SIs junction}

In the experimentally realized case \cite{Bolginov, Ryazanov3,
Larkin, Vernik} $\Gamma _{BI}^{-1}\ll |\beta |\Gamma ^{-1}$ the condition is fulfilled and
the weak place in SIsFS structure is located at the SIs interface. In this
approximation it follows from (\ref{EqForKhi1}) that
\begin{equation*}
\chi \approx \varphi -\frac{2q\Gamma }{(p-q)\Gamma _{BI}}\sin \left( \varphi
\right)
\end{equation*}%
in 0-state ($d_{F}<d_{F1}$) and
\begin{equation*}
\chi \approx \pi +\varphi -\frac{2q\Gamma }{(p-q)\Gamma _{BI}}\sin \left(
\varphi \right)
\end{equation*}%
in $\pi $-state ($d_{F}>d_{F1}$). Substitution of these expressions into (%
\ref{Jsymm1}) results in
\begin{equation}
J_{S}(\varphi )=\pm \frac{J_{G}}{\Gamma _{BI}}\left[ \sin \varphi -\frac{%
\Gamma }{\Gamma _{BI}}\frac{1\mp \beta }{2\beta }\sin \left( 2\varphi
\right) \right]  \label{SISIotFI3}
\end{equation}%
for $0$- and $\pi $- states, respectively. It is seen that for $d_{F}<d_{F1}$
the CPR (\ref{SISIotFI3}) has typical for SIS tunnel
junctions sinusoidal shape with small correction taking into account the
suppression of superconductivity in the s layer due to proximity with FS
part of complex sFS electrode. Its negative sign is typical for the tunnel
Josephson structures with composite NS or FS electrodes \cite{KuperD, CurPh}%
. For $d_{F}>d_{F1}$ the supercurrent changes its sign thus exhibiting
the transition of SIsFS junction into $\pi $ state. It's important to note that
in this mode the SIsFS structure may have almost the same value of the
critical current both in $0$ and $\pi $ states. It is unique property, which
can not be realized in SFS devices studied before. For this reason we
have identified this mode as "Switchable $0-\pi $ SIS junction".

\subsubsection{Mode (1b): sFS junction}

Another limiting case is realized under the condition $\Gamma _{BI}^{-1}\gg
|\beta |\Gamma ^{-1}.$ It fulfills in the vicinity of the points of $0-$ to $%
\pi -$ transitions, $d_{Fn},$ and for large $d_{F}$ values and high exchange
fields $H.$ In this mode (see Fig. \ref{Phase}) the weak place shifts to sFS
part of SIsFS device and the structure transforms into a conventional
SFS-junction with complex SIs electrode.

In the first approximation on $\Gamma /(\beta \Gamma _{BI})\gg 1$ it follows
from (\ref{Jsymm1}), (\ref{EqForKhi1}) that
\begin{equation*}
\chi =\frac{\Gamma _{BI}}{\Gamma }\frac{\beta \sin \left( \varphi \right) }{%
1-\beta \cos \left( \varphi \right) },
\end{equation*}%
resulting in
\begin{equation}
J_{S}(\varphi )=\frac{J_{G}\beta }{\Gamma \left( 1-\beta \cos \varphi
\right) }\left( \sin \varphi -\frac{\Gamma _{BI}}{2\Gamma }\frac{\beta \sin
(2\varphi )}{\left( 1-\beta \cos \varphi \right) }\right) .
\label{SInFSotFI}
\end{equation}%
The shape of CPR for $\chi \rightarrow 0$
coincides with that previously found in SNS and SFS Josephson devices \cite%
{Ivanov}. It transforms to the sinusoidal form for sufficiently large
thickness of F layer. For small thickness of the F-layer as well as in the
vicinity of $0-\pi $ transitions, significant deviations from sinusoidal
form may occurred.

Transition between the \emph{mode (1a)} and the \emph{mode (1b)} is also
demonstrated in Fig.\ref{Ic_ds}. It shows dependence of critical current $%
J_{C}$ across the SIsFS structure versus F-layer thickness $d_{F}.$ The
inset in Fig.\ref{Ic_ds} demonstrates the magnitude of an order parameter at Is
interface as a function of $d_{F}$. The solid lines in Fig.\ref{Ic_ds} give
the shape of $J_{C}(d_{F})$ and $\delta _{0}(-d_{s})$ calculated from
(\ref{SISIotFI3})-(\ref{SInFSotFI}). These equations are valid in the limit $%
d_{s}\gg d_{sc}$ and do not take into account possible suppression of
superconductivity in a vicinity of tunnel barrier due to proximity with FS
part of the device. The dashed lines are the result of calculations using
analytical expressions (\ref{JatIS1})-(\ref{EqDsmall1}) for the thickness of
the s-layer $d_{s}=2\xi _{s}(T),$ which slightly exceeds the critical one, $%
d_{sc}=(\pi /2)\xi _{s}(T).$ These analytical dependencies are calculated at
$T=0.9~T_{C}$ for $H=10\pi T_{C},$ $\Gamma _{BI}=200,~\Gamma =5,$ $\gamma
_{B}=0.$ The short-dashed curves are the results of numerical calculations
performed selfconsistently in the frame of the Usadel equations (2)-(11) for
corresponding set of the parameters $T=0.9~T_{C}$ for $H=10\pi T_{C},$ $%
\gamma _{BI}=1000,$ $\gamma =1,$ $\gamma _{B}=0.3$ and the same thickness of
the s layer $d_{sc}=(\pi /2)\xi _{s}(T)$. Interface parameters $\gamma
_{BI}=1000,$ $\gamma =1$ are chosen the same as for the analytical case. The
choice of $\gamma _{B}=0.3$ allows one to take into account the influence of
mismatch which generally occurs at the sF and FS boundaries.

\begin{figure*}[tbp]
\center{\includegraphics[width=18 cm]{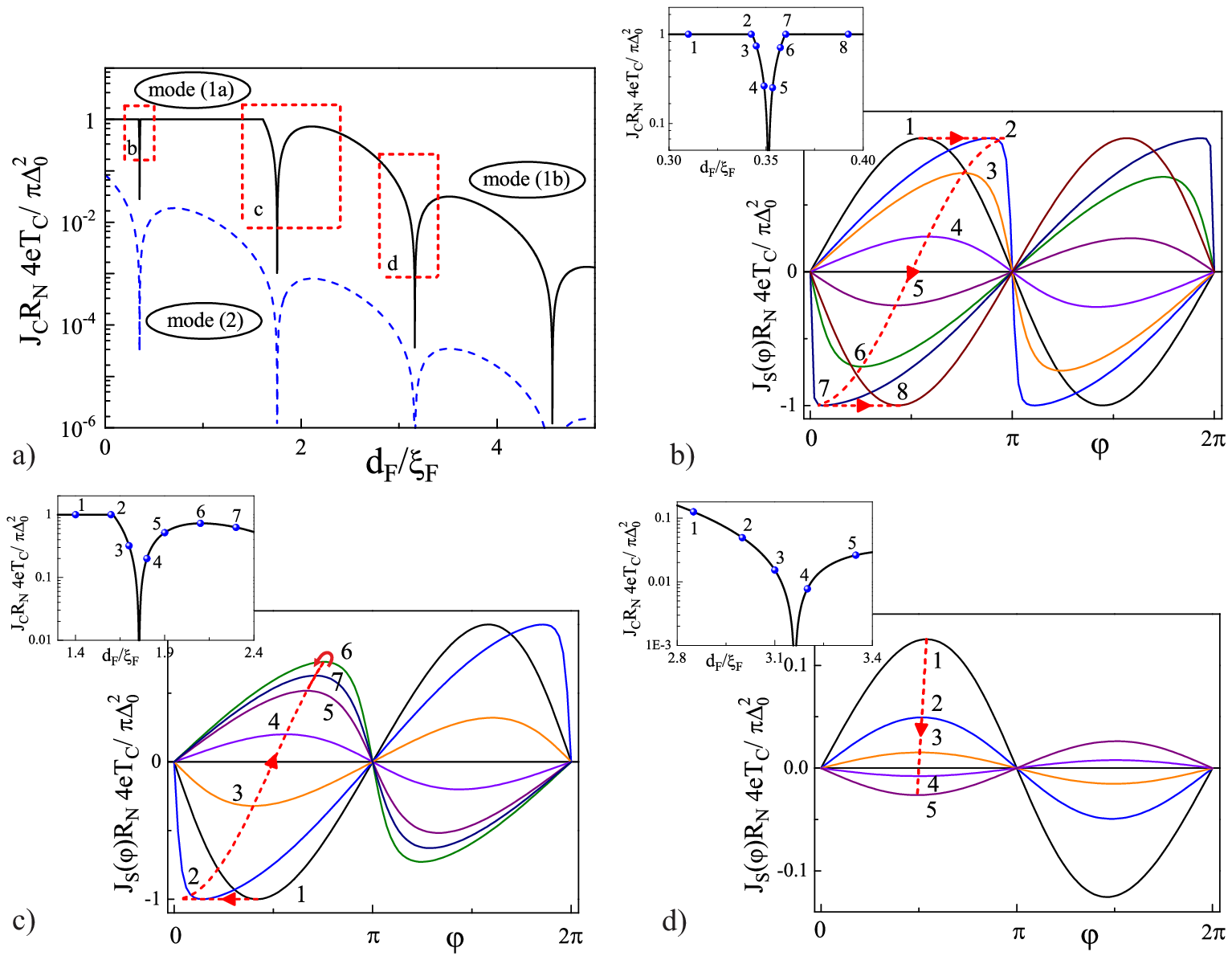}}
\caption{a) Magnitude of the critical current $J_{C}$ in the SIsFS structure
versus F-layer thickness $d_{F}$ for two thickness of middle s-layer, $%
d_{s}=5\protect\xi _{S}(T)>d_{sc}$ (solid line) and $d_{s}=0.5\protect\xi %
_{S}(T)<d_{sc}$ (dashed line) calculated at $T=0.9T_{C}$ for $H=10\protect%
\pi T_{C}$, $\Gamma _{BI}=200$ and $\Gamma =5$. b)-d) CPR in the vicinity of $0$-$\protect\pi $
transitions. The corresponding insets show the enlarged parts of $J_{C}(d_{F})$
dependence enclosed in rectangles on the part a) of the Figure and marked by
the letters b-d, respectively. The digits on the insets show the points at which
the $J_{S}(\protect\varphi )$ curves have been calculated. The dashed lines
in the Figs.\protect\ref{CPR}b-d are the loci of critical points at which
the $J_{S}(\protect\varphi )$ dependence reaches its maximum value $%
J_{C}(d_{F})$. }
\label{CPR}
\end{figure*}

It can be seen that there is a qualitative agreement between the shapes of
the three curves. For small $d_{F}$ the structure is in the $0$-state \emph{%
mode (1a)} regime. The difference between dashed and short dashed lines in
this area is due to the fact that the inequalities (\ref{condSup}) are not
fulfilled for very small $d_{F}.$ The solid and short dashed curves start
from the same value since for $d_{F}=0$ the sFS electrode becomes a single
spatially homogeneous superconductor. For $d_{s}=2\xi _{s}(T)$ the intrinsic
superconductivity in the s layer is weak and is partially suppressed with $d_{F}$
increase (see the inset in Fig.\ref{Ic_ds}). This suppression is accompanied by rapid
drop of the critical current. It can be seen that starting from the value $%
d_{F}\approx 0.4\xi _{F}${}{} our analytical formulas (\ref{JatIS1})-(\ref%
{EqDsmall1}) are accurate enough. The larger $d_{s},$ the better agreement
between numerical and analytical results due to the better applicability of
the GL equations in the s layer. With further $d_{F}$ increase the structure
passes through the valley of \emph{mode (1b)} state, located in the vicinity
of the $0$ to $\pi $ transition, and comes into the $\pi -$state of the
\emph{mode (1a)}. Finally for $d_{F}\gtrsim 1.6\xi _{F}$ there is a
transition from \emph{mode (1a)} to \emph{mode (1b),} which is accompanied
by damped oscillation of $J_{C}(d_{F})$ with $d_{F}$ increase.

\subsection{Mode (2): SInFS junction $d_{s}\leq d_{sc}$}

For $d_{s}\leq d_{sc}$ intrinsic superconductivity in the $s$ layer is
completely suppressed resulting in formation of the complex -InF- weak link
area, where 'n' marks the intermediate s film in the normal state. In this
parameter range the weak is always located in the tunnel barrier and the CPR
has sinusoidal shape
\begin{equation}
J_{S}(\varphi )=\frac{J_{G}}{\sqrt{2}}\frac{\left( p-q\right) \sin \varphi }{%
2pq\Gamma \Gamma _{BI}\cos \frac{d_{s}}{\xi _{s}(T)}+\left[ 2pq\Gamma
+\left( p+q\right) \Gamma _{BI}\right] \sin \frac{d_{s}}{\xi _{s}(T)}}.
\label{I_ot_fi_small_ds1}
\end{equation}%
In a vicinity of the critical thickness, $d_{s}\lesssim d_{sc},$ the factor $%
\cos (d_{s}/\xi _{S}(T))$ in (\ref{I_ot_fi_small_ds1}) is small and
supercurrent is given by the expression
\begin{equation}
J_{S}(\varphi )=\frac{J_{G}}{2\sqrt{2}}\frac{\left( p-q\right) \sin \varphi
}{2pq\Gamma +\left( p+q\right) \Gamma _{BI}}.  \label{curInter}
\end{equation}
Further decrease of $d_{s}$ into the limit $d_{s}\ll d_{sc}$ leads to
\begin{equation}
J_{S}(\varphi )=\frac{J_{G}}{\sqrt{2}}\frac{\left( p-q\right) \sin \varphi }{%
2pq\Gamma \Gamma _{BI}}.  \label{curSInFS}
\end{equation}%
The magnitude of critical current in (\ref{curSInFS}) is close to that in the
well-known case of SIFS junctions in appropriate regime.

\subsection{Current-Phase Relation \label{SecCPR}}

In the previous section we have demonstrated that the variation in the
thickness of the ferromagnetic layer should lead to the transformation of
CPR of the SIsFS structure. Fig.\ref{CPR}a illustrates the $J_{C}(d_{F})$
dependencies calculated from expressions (\ref{JatIS1})-(\ref{EqDsmall1}) at
$T=0.9T_{C}$ for $H=10\pi T_{C},$ $\gamma _{B}=0,$ $\Gamma _{BI}\approx 200$
and $\Gamma \approx 5$ for two thickness of the s layer $d_{s}=5\xi _{S}(T)$
(solid line) and $d_{s}=0.5\xi _{S}(T)$ (dashed line). In Figs.\ref{CPR}b-d
we enlarge the parts of $J_{C}(d_{F})$ dependence enclosed in rectangles
labeled by letters b, c and d in Fig.\ref{CPR}a and mark by digits
the points where the $J_{S}(\varphi )$ curves have been calculated. These
curves are marked by the same digits as the points in the enlarge parts of $%
J_{C}(d_{F})$ dependencies. The dashed lines in the Figs.\ref{CPR}b-d are
the loci of critical points at which the $J_{S}(\varphi )$ dependence
reaches its maximum value, $J_{C}(d_{F})$.

Figure \ref{CPR}b presents the \emph{mode (1b)} valley, which divides the
\emph{mode (1a)} domain into $0$- and $\pi $- states regions. In the \emph{%
mode (1a)} domain the SIsFS structure behaves as SIs and sFS junctions
connected in series. Its critical current equals to the minimal one among
the critical currents of the SIs $(J_{CSIs})$ and sFS $(J_{CsFS})$ parts of
the device. In the considered case the thickness of the s film is
sufficiently large to prevent suppression of superconductivity.  Therefore, $J_{CSIs}$
does not change when moving from the point $1$ to the point $2$ along $J_{C}d_{F}$
dependence. At the point $2$, when $J_{CSIs}=J_{CsFS},$ we arrive at the border between the \emph{mode (1a) }%
and\emph{\ mode (1b)}. It is seen that at this point there is maximum
deviation of $J_{S}(\varphi )$ from the sinusoidal shape. Further increase of $%
d_{F}$ leads to $0$-$\pi $ transition, when parameter $\beta $ in (\ref%
{SInFSotFI}) becomes small and $J_{S}(\varphi )$ practically restores its
sinusoidal shape. Beyond the area of $0$ to $\pi $ transition, the critical
current changes its sign and CPR starts to deform again.
The deformation achieves its maximum at the point $7$ located at the other
border between the \emph{modes (1a) }and\emph{\ (1b).} The displacement from
the point 7 to the point 8 along the $J_{C}(d_{F})$ dependence leads to recovery of
sinusoidal CPR.

Figure \ref{CPR}c presents the transition from the $\pi $-state of \emph{%
mode (1a)} to \emph{mode (1b)} with $d_{F}$ increase. It is seen that the
offset from the point $1$ to the points $2-5$ along $J_{C}(d_{F})$ results
in transformation of the CPR similar to that shown in Fig.\ref{CPR}b during displacement in the direction from the
point $1$ to the points $2-6.$ The only difference is the starting negative
sign of the critical current. However this behavior of CPR as well as close
transition between modes lead to formation of the well pronounced kink at the $%
J_{C}(d_{F})$ dependence. Furthermore, contrary to Fig.\ref{CPR}b at the
point $6$, the junction is still in the \emph{mode (1b) } and remains in this
mode with further increase in $d_{F}.$ At the point $6$ the critical current
achieves its maximum value and it decreases along the dashed line for larger
$d_{F}.$

Figure \ref{CPR}d \ shows the transformation of the CPR in the
vicinity of the next $0$ to $\pi $ transition in  \emph{mode (1b). } There is
small deviation from sinusoidal shape at the point $1$, which vanishes
exponentially with an increase of \thinspace $d_{F}$.

In the\emph{\ mode (2)} (the dashed curve in Fig.\ref{CPR}a) an intrinsic
superconductivity in the s layer is completely suppressed resulting in the
formation of a complex -InF- weak link region and the CPR becomes sinusoidal (\ref{I_ot_fi_small_ds1}).

\section{Arbitrary temperature}

\begin{figure*}[tbp]
\center{\includegraphics[width=16 cm]{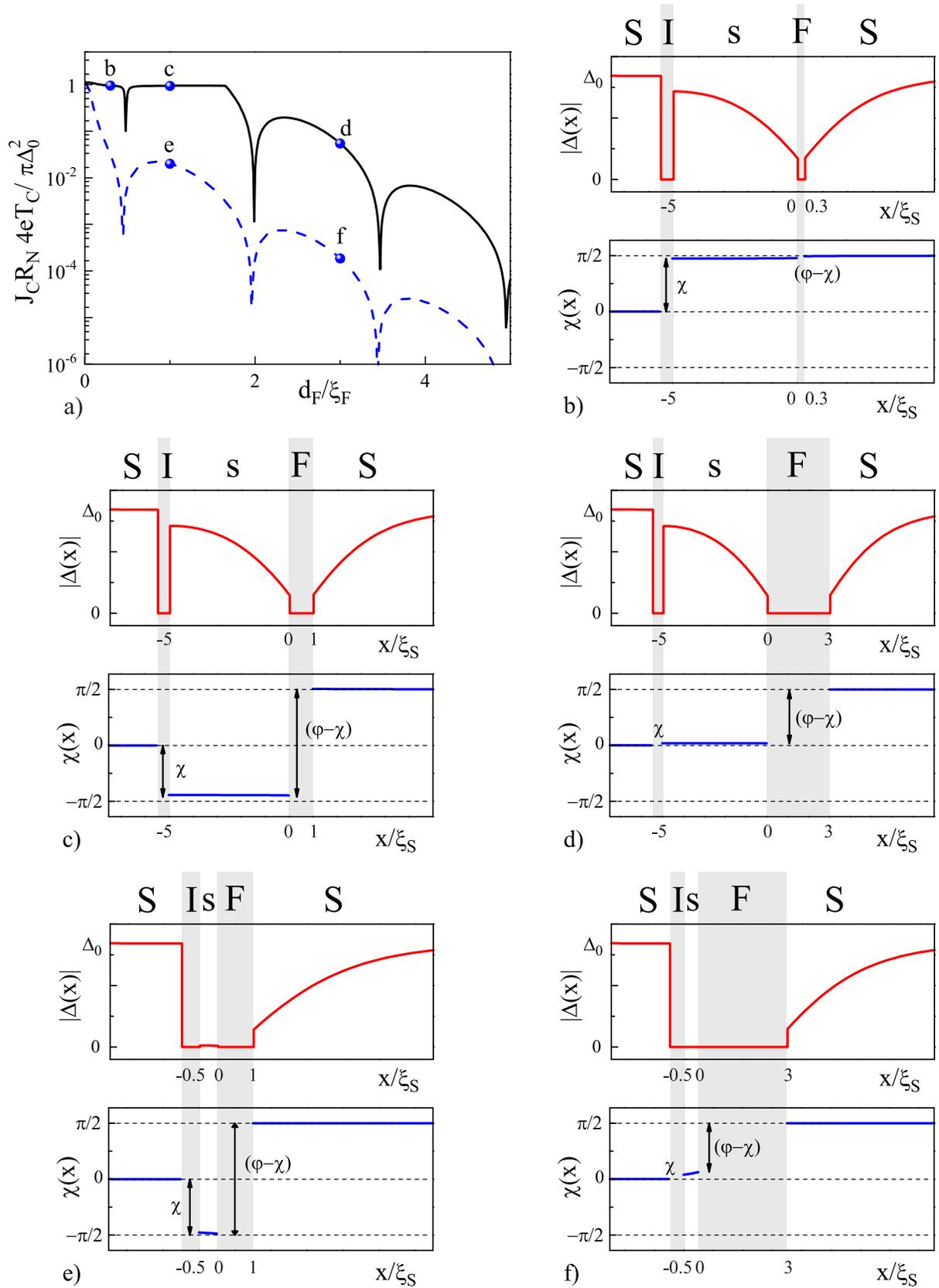}}
\caption{a) Magnitude of critical current $J_{C}$ of the SIsFS structure
versus F-layer thickness $d_{F}$ calculated at $T=0.5T_{C}$ for $H=10\protect%
\pi T_{C}$, $\protect\gamma _{BI}=1000,$ $\protect\gamma =1$ and two
thickness of the s film $d_{s}=5\protect\xi _{S}$ (solid line) and $d_{s}=0.5%
\protect\xi _{S}$ (dashed line). The letters on $J_{C}(d_{F})$ give the
points at which the coordinate dependencies of the magnitude of the order
parameter, $|\Delta (x)|,$ and phase difference across the structure, $%
\protect\chi ,$ have been calculated. These curves are shown in the panels
b)-f) of the Figure as the upper and bottom panels, respectively. }
\label{Del_ArbT}
\end{figure*}
At arbitrary temperatures the boundary problem (\ref{fiS})-(\ref{PhDifInf})
goes beyond the assumptions of GL formalism and requires self-consistent
solution. We have performed it numerically in terms of the nonlinear Usadel
equations in iterative manner. All calculations were performed for $%
T=0.5T_{C},$ $\xi _{S}=\xi _{F},$ $\gamma _{BI}=1000,$ $\gamma _{BFS}=0.3$
and $\gamma =1.$

Calculations show that at the selected transparency of
tunnel barrier $(\gamma _{BI}=1000)$ the suppression of superconductivity in the
left electrode is negligibly small. This allows one to select the thickness of the
left S electrode $d_{SL}=2\xi _{S}$ without any loss of generality. On the
contrary, proximity of the right S electrode to the F layer results in
strong suppression of superconductivity at the FS interface. Therefore the pair
potential of the right S electrode reaches its bulk value only at thickness $%
d_{SR}\gtrsim 10\xi _{S}.$ It is for these reasons we have chosen $d_{SR}=10\xi _{S}$
for the calculations.

Furthermore, the presence of a low-transparent tunnel barrier in the
considered SIsFS structures limits the magnitude of
critical current $J_{C}$ by a value much smaller compared to a
depairing current of the superconducting electrodes. This allows one to neglect
nonlinear corrections to coordinate dependence of the phase
in the S banks.

The results of calculations are summarized in Fig.\ref{Del_ArbT}. Figure \ref%
{Del_ArbT}a shows the dependence of $J_{C}$ of the
SIsFS structure on the F-layer thickness $d_{F}$ for relatively large $%
d_{s}=5\xi _{S}$ (solid) and small $d_{s}=0.5\xi _{S}$ (dashed) s-film
thickness. The letters on the curves indicate the points at which the coordinate
dependencies of the magnitude of the order parameter, $|\Delta (x)|,$ and
phase difference across the structure, $\chi ,$ have been calculated for the
phase difference $\varphi = \pi /2$. These curves are shown in the panels b)-f) of the
Fig.\ref{Del_ArbT} as the upper and bottom plots, respectively. There is
direct correspondence between the letters, b, c, d, e, f, on $J_{C}(d_{F})$
curves and the labels, b), c), d), e), f), of the panels.

It is seen that qualitative behavior of the $J_{C}(d_{F})$ dependence at $%
T=0.5T_{C}$ remains similar to that obtained in the frame of the GL equations
for $T=0.9T_{C}$ (see Fig.\ref{CPR}a). Furthermore, the modes of operation
discussed above remain relevant too. The panels b)-f) in Fig.\ref{Del_ArbT}
make this statement more clear.

At the point marked by letter 'b', the s-film is sufficiently thick, $d_{s}=5\xi
_{S},$ while F film is rather thin, $d_{F}=0.3\xi _{F},$ and therefore the
structure is in $0$- state of the \emph{mode (1a)}. In this regime the
phase mainly drops across the tunnel barrier, while the phase
shifts at the s-film and in the S electrodes are negligibly small(see the
bottom plot in Fig.\ref{Del_ArbT}b).

At the point marked by the letter 'c' $(d_{s}=5\xi _{S},$ $d_{F}=\xi _{F})$, the
structure is in the $\pi $- state of the \emph{mode (1a)}. It is seen from
Fig.\ref{Del_ArbT}c that there is a phase jump at the tunnel barrier and
an additional $\pi $-shift occurs between the phases of S and s layers.

For $d_{F}=3\xi _{F}$ (Fig.\ref{Del_ArbT}d) the position of the weak place
shifts from SIs to sFS part of the SIsFS junction. Then the structure starts to operate in
the \emph{mode (1b)}. It is seen that the phase drop across SIs  part is
small, while $\varphi -\chi \approx $ $\pi /2$ across the F layer, as it
should be in SFS junctions with SIs and S electrodes.

At the points marked by the letters 'e' and 'f', thickness of the s-layer $%
d_{s}=0.5\xi _{S}$ is less than its critical value. Then superconductivity in the
s-spacer is suppressed due to the proximity with the F film and SIsFS device
operates in the \emph{mode (2)}. At $d_{F}=\xi _{F}$ (the dot 'e' in Fig.\ref%
{Del_ArbT}a and the panel Fig.\ref{Del_ArbT}e) the position of the weak
place is located at the SIs part of the structure and there is additional $\pi $%
-shift of phase across the F film. As a result, the SIsFS structure
behaves like an SInFS tunnel $\pi $-junction. Unsuppressed residual value of
the pair potential is due to the proximity with the right S-electrode and it
disappears with the growth of the F-layer thickness, which weakens this proximity
effect. At $d_{F}=3\xi _{F}$ (Fig.\ref{Del_ArbT}f) weak place is located at
the F part of IsF trilayer. Despite strong suppression of the pair
potential in the s-layer, the distribution of the phase inside the IsF
weak place has rather complex structure, which depends on thicknesses of the s
and F layers.

\subsection{Temperature crossover from 0 to $\protect\pi $ states}

The temperature-induced crossover from 0 to $\pi $ states in SFS\
junctions has been discovered in \cite{ryazanov2001} in
structures with sinusoidal CPR.
It was found that the transition takes place in a relatively broad temperature range.

\begin{figure*}[tbh]
\begin{minipage}[h]{0.49\linewidth}
\center{\includegraphics[width=1\linewidth]{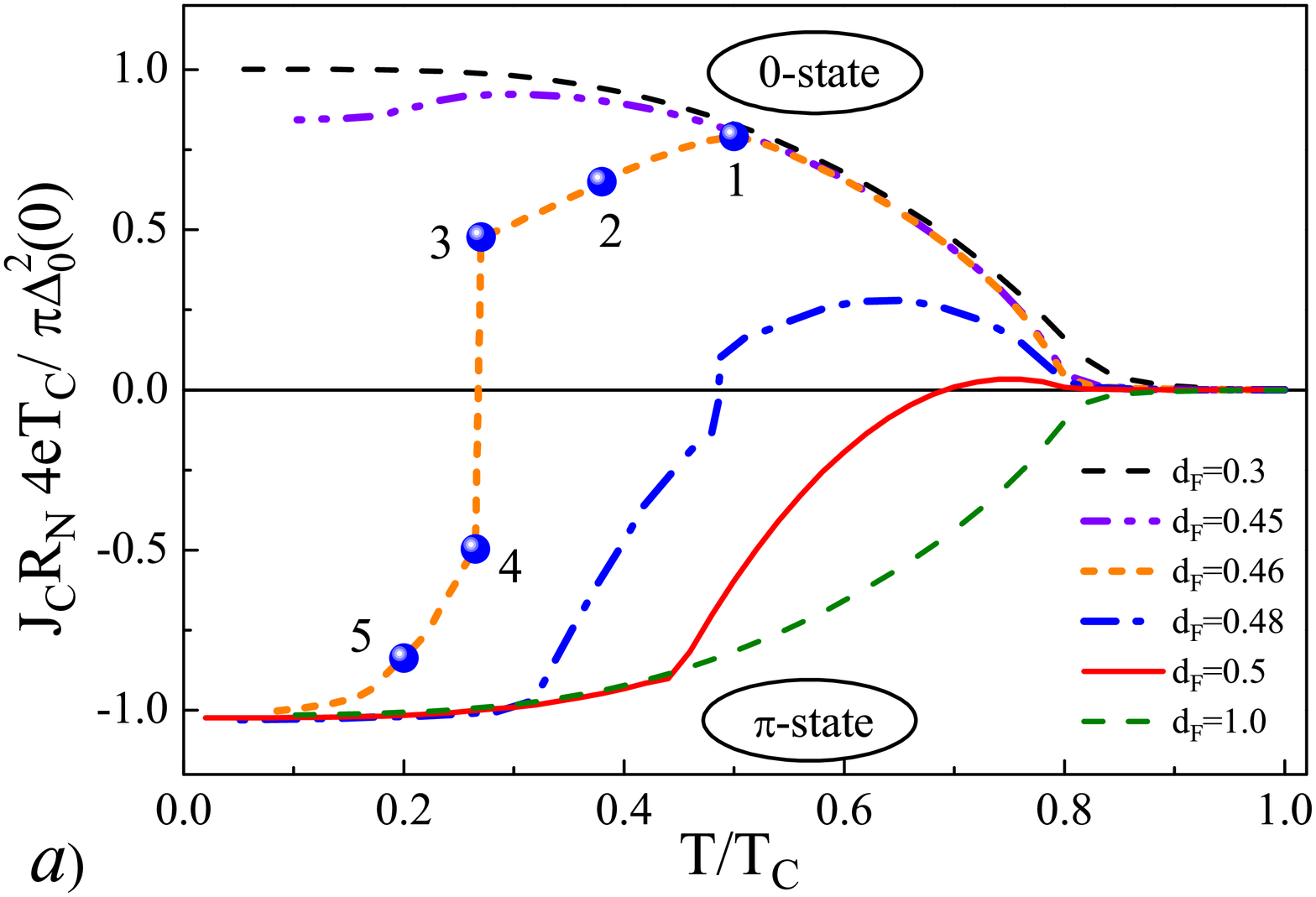}}
\end{minipage}
\hfill
\begin{minipage}[h]{0.49\linewidth}
\center{\includegraphics[width=1\linewidth]{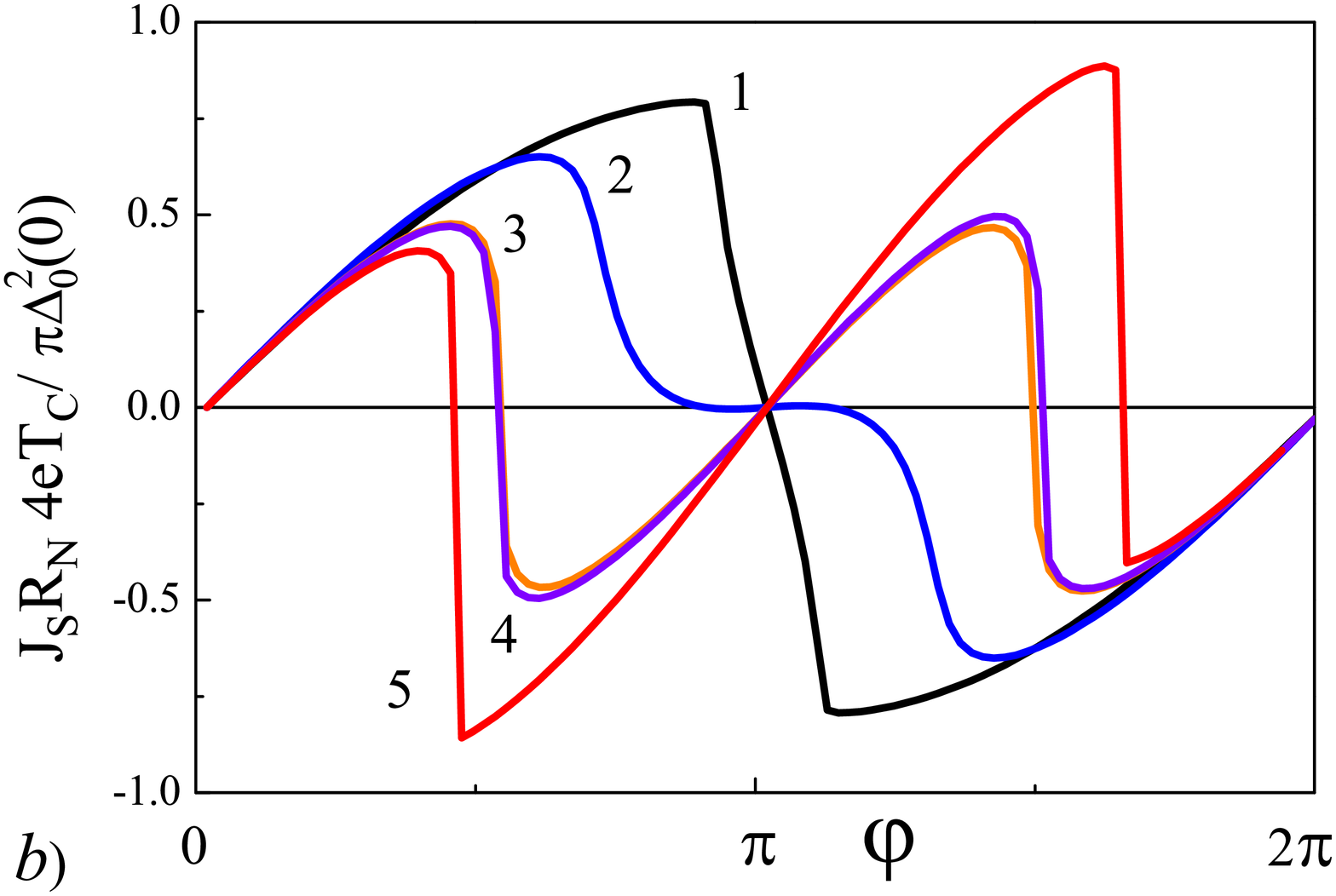}}
\end{minipage}
\caption{a) Critical current $J_{C}$ of the SIsFS structure versus
temperature $T$ for various F-layer thicknesses $d_{F}$ in the vicinity
of $0$ to $\protect\pi $ transition. Dashed envelopes show
temperature dependence in the $0$- (top) and $\protect\pi $- (bottom)
states. b) CPR of the structure with $d_{F}=0.46\protect%
\xi _{F}$ for a set of the temperatures in the vicinity of $0$- $\protect\pi $
transition. Each curve corresponds to the point marked in the panel (a). Note
that the curves (3) and (4) almost coincide but correspond to
different ground states of the junction, $0$- and $\protect\pi $, respectively. The calculations have
been performed numerically for $d_{s}=5\protect\xi _{S}$, $H=10\protect\pi %
T_{C}$, $\protect\gamma _{BI}=1000,$ $\protect\gamma _{B}=0.3,$ $\protect%
\gamma =1$. }
\label{TempTrans}
\end{figure*}

Our analysis of SIsFS structure (see Fig.\ref{TempTrans}a)
shows that smoothness of 0 to $\pi $ transition strongly depends on the CPR
shape. This phenomenon was not analyzed before since almost all previous theoretical results were obtained
within a linear approximation leading in a sinusoidal CPR. To prove the statement, we
have calculated numerically the set of $J_{C}(T)$ curves for a number of F layer
film thicknesses $d_{F}.$ We have chosen the thickness of intermediate
superconductor $d_{S}=5\xi _{S}$ in order to have SIsFS device in
the \emph{mode(1a)} and we have examined the parameter range $0.3\xi _{F}\leq d_{F}\leq \xi _{F},$
in which the structure exhibits the first $0$ to $\pi $ transition. The
borders of the $d_{F}$ range are chosen in such a way that SIsFS contact is either
in $0$- ($d_{F}=0.3\xi _{F}$) or $\pi $- ($d_{F}=\xi _{F}$) state in the
whole temperature range. The corresponding $J_{C}(T)$ dependencies (dashed
lines in Fig. \ref{TempTrans}a) provide the envelope of a set of $J_{C}(T)$
curves calculated for the considered range of $d_{F}$. It is
clearly seen that in the vicinity of $T_{C}$ the decrease of $d_{F}$ results
in creation of the temperature range where $0$-state exists. The point of $0$
to $\pi $ transition shifts to lower temperatures with decreasing $d_{F}$.
For $d_{F}\gtrsim 0.5\xi _{F}$ the transition is rather smooth
since for $T\geq 0.8T_{C}$ the junction keeps the \emph{mode (2)} (with suppressed
superconductivity) and deviations of the CPR from $\sin (\varphi )$
are small. Thus the behavior of $J_{C}(T)$ dependencies in this case can be easily
described by analytic results from Sec.\ref{SecCPR}.

The situation drastically changes at $d_{F}=0.46\xi _{F}$ (short-dashed
line in Fig.\ref{TempTrans}a). For this thickness the point of $0$ to $\pi $
transition shifts to $T\approx 0.25T_{C}.$ This shift is accompanied by an
increase of amplitudes of higher harmonics of CPR (see Fig.\ref{TempTrans}b).
As a result, the shape of CPR is strongly modified, so that in the interval $0\leq $ $\varphi \leq $ $\pi $
the CPR curves are characterized by two values, $J_{C1}$ and $J_{C2}$, as is known from the
case of SFcFS constrictions \cite{SFcFS}. In general, $J_{C1}$ and $J_{C2}$
differ both in sign and magnitude and $J_{C}=\max (\left\vert
J_{C1}\right\vert ,$ $\left\vert J_{C2}\right\vert ).$ For $T>0.25T_{C}$ the
junction in the $0$-state and $J_{C}$ grows with decrease of $T$ up to $%
T\approx 0.5T_{C}.$ Further decrease of $T$ is accompanied by suppression of
critical current. In a vicinity of $T\approx 0.25T_{C}$ the difference
between $\left\vert J_{C1}\right\vert ,$ and $\left\vert J_{C2}\right\vert $
becomes negligible and the system starts to develop the instability that
eventually shows up as a sharp jump from $0$ to $\pi $ state. After the
jump, $\left\vert J_{C}\right\vert $ continuously increases when $T$
goes to zero.

It is important to note that this behavior should always be observed in the
vicinity of $0-\pi $ transition, i.e. in the range of parameters, in which
the amplitude of the first harmonic is small compared to higher
harmonics. However, the closer is temperature to $T_c$, the less pronounced are higher CPR harmonics
and the smaller is the magnitude of the jump. This fact is illustrated by dash-dotted line showing
$J_{C}(T)$ calculated for $d_{F}=0.48\xi _{F}.$ The jump in the curves
calculated for $d_{F}\geq 0.5\xi _{F}$ also exists, but it is small and can not be
resolved on the scale used in the Fig. \ref{TempTrans}a.

At $d_{F}=0.45\xi _{F}$ (dash-dot-dotted line in Fig. \ref{TempTrans}) the
junction is always in the $0$-state and there is only small suppression of
critical current at low temperatures despite the realization of non-sinusoidal CPR.

Thus the calculations clearly show that it's possible to realize a set of
parameters of SIsFS junctions where thermally-induced $0$-$\pi $ crossover can be 
observed and controlled by temperature variation.
\begin{figure}[t]
\center{\includegraphics[width=1\linewidth]{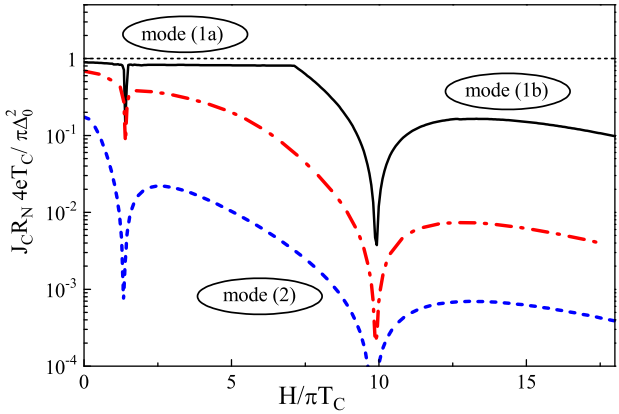}}
\caption{The magnitude of critical current $J_{C}$ of the SIsFS structure versus exchange
field $H$ for thick $d_{s}=5\protect\xi _{S}$ (solid), thin $d_{s}=0.5%
\protect\xi _{S}$ (dashed) and intermediate $d_{s}=3\protect\xi _{S}$
(dash-dotted) s-layer thickness. The plot demonstrates the possibility of $0$-$%
\protect\pi $ transition by varying the effective exchange field. The
calculations have been performed for $T=0.5T_{C}$, $d_{F}=2\protect\xi _{F}$%
, $\protect\gamma _{BI}=1000,$ $\protect\gamma _{B}=0.3,$ $\protect\gamma =1$.}
\label{JC_H}
\end{figure}

\subsection{0 to $\protect\pi $ crossover by changing the
effective exchange energy in external magnetic field}

Exchange field is an intrinsic microscopic parameter of a
ferromagnetic material which cannot be controlled directly by application of
an external field. However, the spin splitting in F-layers can
be provided by both the internal exchange field and external magnetic field%
\cite{Tedrow, Koshina}, resulting in generation of effective exchange field,
which equals to their sum. However, practical realization of this
effect is a challenge since it is difficult to fulfill special
requirements\cite{Tedrow, Koshina} on thickness of S electrodes and SFS
junction geometry.

Another opportunity can be realized in soft diluted ferromagnetic alloys
like Fe$_{0.01}$Pd$_{0.99}$. Investigations of magnetic properties\cite%
{Uspenskaya} of these materials have shown that below 14 K they exhibit
ferromagnetic order due to the formation of weakly coupled
ferromagnetic nanoclusters. In the clusters, the effective spin polarization
of Fe ions is about $4\mu _{B}$, corresponding to that in the bulk Pd$_{3}$%
Fe alloy. It was demonstrated that the hysteresis loops of Fe$_{0.01}$Pd$%
_{0.99}$ films have the form typical to nanostructured ferromagnets with
weakly coupled grains (the absence of domains; a small coercive force; a
small interval of the magnetization reversal, where the magnetization changes its direction
following the changes in the applied magnetic field; and a prolonged part, where the component of
the magnetization vector along the applied field grows gradually).

Smallness of concentration of Pd$_{3}$Fe clusters and their ability to follow variation
in the applied magnetic field may result in generation of $H_{eff},$ which
is of the order of
\begin{equation}
H_{eff}\approx H\frac{n_{\uparrow }V_{\uparrow }-n_{\downarrow
}V_{\downarrow }}{n_{\downarrow }V_{\downarrow }+n_{\uparrow }V_{\uparrow
}+(n-n_{\uparrow }-n_{\downarrow })(V-V_{\downarrow }-V_{\uparrow })}.
\label{Heff}
\end{equation}%
Here $n$ is concentration of electrons within a physically small volume $%
V $, in which one performs an averaging of Greens functions in the
transformation to a quasiclassical description of
superconductivity, $n_{\uparrow ,\downarrow }$ and $V_{\uparrow ,\downarrow
} $ are the values describing spin polarized parts of $n$ and parts
of volume $V,$ which they occupy, respectively. Similar kind of $H_{eff}$
nucleates in NF or SF proximity structures, which are composed from thin
layers \cite{BerVE, BChG, Karminskaya, Golikova}. There is an interval of
applied magnetic fields $H_{ext}$ where the alloy magnetization changes its direction
and the concentrations $n_{\uparrow ,\downarrow }$
depend on a pre-history of application of the field \cite{Larkin, APL}, 
providing the possibility to control $H_{eff}$ by an external magnetic field.

Derivation of possible relationships between $H_{eff}$ and $H_{ext}$ is
outside of the scope of this paper. Below we will concentrate only on an
assessment of the intervals in which $H_{eff}$ should be changed to ensure
the transition of SIsFS device from $0$ to $\pi $ state. To do this, we
calculate the $J_{C}(H)$ dependencies presented in Fig.\ref{JC_H}. The
calculations have been done for the set of structures with $d_{F}=2\xi _{F}$
and s-films thickness ranging from thick one, $d_{S}=5\xi _{S}$ (solid line)
up to an intermediate value $d_{S}=2\xi _{S}$ (dashed-dotted line) and
finishing with thin film having $d_{S}=0.5\xi _{S}$ (dashed line). It is clearly seen that
these curves have the same shape as $J_{C}(d_{F})$ dependencies presented in
the Sec.\ref{SecHighTC}. For $d_{S}=5\xi _{S}$ and $%
H\lesssim 7\pi T_{C}$ the magnitude of $J_{C}$ is practically independent on
$H,$ but it changes the sign at $H\approx 1.25\pi T_{C}$ due to $0$ - $\pi $
transition. It is seen that for the transition, while maintaining the
normalized current value at a level close to unity, changes of $H$ are required
approximately of the order of $0.1\pi T_{C}$ or $10\%$. For $d_{S}=2\xi _{S}$
and $H\lesssim 3\pi T_{C}$, it is necessary to change $H$ on $20\%$
to realize the such a transition. In this case the value
of normalized current is at the level $0.4$. In mode 2 the
transition requires $100\%$ change of $H,$ which is not practical.

\section{Discussion}

We have performed a theoretical study of magnetic SIsFS Josephson junctions.
At $T\leq T_{C}$ calculations have been performed analytically in the frame
of the GL equations. For arbitrary temperatures we have developed
numerical code for selfconsistent solution of the Usadel equations. We have
outlined several modes of operation of these junctions. For s-layer in
superconducting state they are S-I-sfS or SIs-F-S devices with weak place
located at insulator (\emph{mode (1a)}) and at the F-layer (\emph{mode (1b)}%
), respectively. For small s-layer thickness, intrinsic superconductivity in
it is completely suppressed resulting in formation of InF weak place (\emph{%
mode (2)}). We have examined the shape of $J_{S}(\varphi )$ and spatial
distribution of the module of the pair potential and its phase difference
across the SIsFS structure in these modes.

For \emph{\ mode (1) } the shape of the CPR can substantially differ from the sinusoidal one
even in a vicinity of $T_{C}$.
The deviations are largest when the structure is close to the crossover between the \emph{modes
(1a)} and\emph{\ (1b)}. This effect results in the kinks in the
dependencies of $J_{C}$ on temperature and on parameters of the structure
(thickness of the layers $d_{F},$ $d_{s}$ and exchange energy $H$) as
illustrated in Fig.\ref{CPR} on $J_{C}(d_{F})$ curves. The transformation
of CPR is even more important at low temperatures. For
$T\lesssim 0.25T_{C}$ a sharp $0-\pi $ transition can be realized induced
by small temperature variation (Fig.\ref{TempTrans}). This instability must
be taken into account when using the structures as memory elements. On the
other hand, this effect can be used in detectors of electromagnetic radiation,
where absorption of a photon in the F layer will provide local heating
leading to development of the instability and subsequent phonon registration.

We have shown that suppression of the order parameter in the thin s-film due
to the proximity effect leads to decrease of $J_{C}R_{N}$ product in both $0-$
and $\pi -$states. On the other hand, the proximity effect may also support s-layer
superconductivity due to the impact of S electrodes. In \emph{mode (1a) } $%
J_{C}R_{N}$ product in $0$- and $\pi $-states can achieve values typical for
SIS tunnel junctions.

In \emph{mode (2) } sinusoidal CPR is realized. Despite that, the distribution
of the phase difference $\chi (x)$ in the IsF weak place may have a complex
structure, which depends on thickness of the s and F layers. These effects
should influence the dynamics of a junction in its \textit{ac}-state and deserve further
study.

Further, we have also shown that in \emph{\ mode (1a)} nearly $10\%$ change
in the exchange energy can cause a $0-\pi $ transition, i.e. changing the
sign of $J_{C}R_{N}$ product, while maintaining its absolute value. This
unique feature can be implemented in \emph{\ mode (1a)}, since it is in it
changes of the exchange energy only determine the presence or absence of a $%
\pi$ shift between s and S electrodes and does not affect the magnitude of
the critical current of SIs part of SIsFS junction.

In \emph{\ mode (1b)}, the F layer becomes a part of weak link area. In this
case the $\pi $ shift, initiated by the change in $H$ must be accompanied by
changes of $J_{C}$ magnitude due to the oscillatory nature of
superconducting correlations in the F film. The latter may lead to
very complex and irregular dependence of $J_{C}(H_{ext}),$ which have been
observed in Nb-PdFe-Nb SFS junctions(see Fig.3 in\cite{Bolginov}). Contrary
to that the $J_{C}(H_{ext})$ curves of SIsFS structure with the same PdFe
metal does not demonstrate these irregularities\cite{Larkin, Vernik}.

\begin{figure}[h]
\center{\includegraphics[width=1\linewidth]{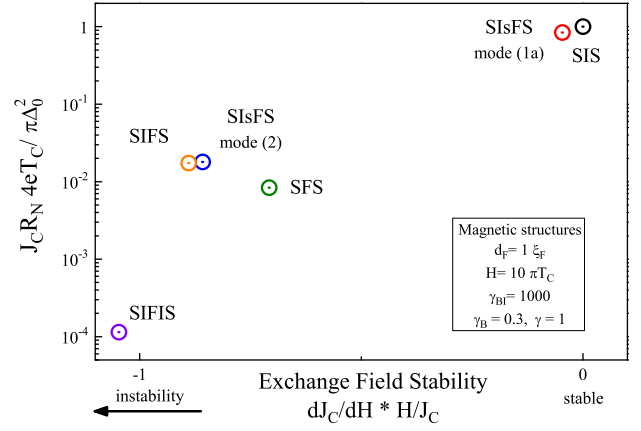}}
\caption{Comparison of different types of Josephson structures, marked by points on the
phase plane, in terms of $J_CR_N$ and exchange field stability $\eta$. 
All calculation have been performed for $T=0.5 T_C$, $d_F=\protect\xi%
_F$, $\protect\gamma _{BI}=1000,$ $\protect\gamma _{B}=0.3,$ $\protect\gamma %
=1$. For SIsFS structures $d_s=5\protect\xi_S$ and $d_s=0.5\protect\xi_S$ 
are taken in \emph{mode (1a)} and \emph{mode (2)}, respectively.}
\label{Compare}
\end{figure}

To characterize a junction stability with respect to $H$ variations it is
convenient to introduce the parameter $\eta =(dJ_{C}/J_{C})/(dH/H)$ which
relates the relative change in the critical current to the relative change
in the exchange energy. The larger
the magnitude of $\eta $ the more intensive irregularities in an SFS
junction are expected with variation of $H$. In the Fig.\ref{Compare} we
compare the SIsFS devices with conventional SFS, SIFS and SIFIS junctions
making use of two the most important parameters: the instability
parameter $\eta $ and $J_{C}R_{N}$ product, the value, which
characterizes high frequency properties of the structures. The calculations
have been done in the frame of Usadel equation for the same set of junctions
parameters, namely $T=0.5 T_C$, $H=10\pi T_{C},$ $d_{F}=\xi _{F},$ $\gamma
_{BI}=1000,$ $\gamma _{B}=0.3,$ $\gamma =1.$ 

It can be seen that the presence of two tunnel barriers in SIFIS junction
results in the smallest $J_{C}R_{N}$ and strong instability.
The SIFS and SIsFS structures in the \emph{mode (2)} demonstrate better results with almost the same parameters.
Conventional SFS structures have two times smaller $J_{C}R_{N} $ product, having
higher critical current but lower resistivity. At the same time, SFS
junctions are more stable due to the lack of low-transparent tunnel barrier.
The latter is the main source of instability due to sharp phase
discontinuities at the barrier 'I'.

Contrary to the standard SFS, SIFS and SIFIS junctions, SIsFS structures
achieve $J_{C}R_{N} $ and stability characteristics comparable to those of SIS tunnel junctions.
This unique property is favorable for application of SIsFS structures in superconducting electronic circuits.

\begin{acknowledgments}
We thank V.V. Ryazanov, V.V. Bol'ginov, I.V. Vernik and O.A. Mukhanov for
useful discussions. This work was supported by the Russian Foundation for
Basic Research, Russian Ministry of Education and Science, Dynasty
Foundation, Scholarship of the President of the Russian Federation, IARPA
and Dutch FOM.
\end{acknowledgments}

\appendix

\section{Boundary problem at $T\lesssim T_{C}$\label{Appendix}}

In the limit of high temperature
\begin{equation}
G_{S}=G_{s}=G_{F}=\sgn(\Omega )  \label{GTTc}
\end{equation}%
and the boundary problem reduces to to the system of linerized equations.
Their solution in the F layer, $(0\leq x\leq d_{F}),$ has the form%
\begin{equation}
\Phi _{F}=C\sinh \frac{\sqrt{\Theta }\left( x-d_{F}/2\right) }{\xi _{F}}%
+D\cosh \frac{\sqrt{\Theta }\left( x-d_{F}/2\right) }{\xi _{F}},
\label{SolinF}
\end{equation}%
where $\Theta =\widetilde{\Omega }\sgn(\Omega ).$ For transparent FS\ and sF
interfaces $(\gamma _{B}=0)$ from the boundary conditions (\ref{BCsF}), (\ref%
{BCSF}) and (\ref{SolinF}) it is easy to get that%
\begin{equation}
\frac{\xi _{s}}{\gamma \sqrt{\Theta }}\frac{d}{dx}\Phi _{s}(0)=-\Phi
_{s}(0)\coth \frac{d_{F}\sqrt{\Theta }}{\xi _{F}}+\frac{\Phi _{S}(d_{F})}{%
\sinh \frac{d_{F}\sqrt{\Theta }}{\xi _{F}}},  \label{Rel1}
\end{equation}%
\begin{equation}
\frac{\xi _{S}}{\gamma \sqrt{\Theta }}\frac{d}{dx}\Phi _{S}(d_{F})=\Phi
_{S}(d_{F})\coth \frac{d_{F}\sqrt{\Theta }}{\xi _{F}}-\frac{\Phi _{s}(0)}{%
\sinh \frac{d_{F}\sqrt{\Theta }}{\xi _{F}}}.  \label{Rel2}
\end{equation}%
and thus reduce the problem to the solution of Ginzburg-Landau (GL)
equations in s and S films.%
\begin{equation}
\xi _{S}^{2}(T)\frac{d^{2}}{dx^{2}}\Delta _{k}-\Delta _{k}(\Delta
_{0}^{2}-\left\vert \Delta _{k}\right\vert ^{2})=0,~\Delta _{0}^{2}=\frac{%
8\pi ^{2}T_{C}(T_{C}-T)}{7\zeta (3)},  \label{GLeq}
\end{equation}%
\begin{equation}
J=\frac{J_{G}}{\Delta _{0}^{2}}\im\left( \Delta _{k}^{^{\ast }}\xi _{S}(T)%
\frac{d}{dx}\Delta _{k}\right) ,~J_{G}=\frac{\pi \Delta _{0}^{2}}{4e\rho
_{S}T_{C}\xi _{S}(T)},  \label{GLcurrent}
\end{equation}%
where $\xi _{S}(T)=\pi \xi _{S}/2\sqrt{1-T/T_{C}}$ is GL coherence length
and $k$ equals to $s$ or $S$ for $-d_{s}\leq x\leq 0$ and $x\geq d_{F\text{ }%
},$ respectively. At Is, sF and FS interfaces GL equations should be
supplemented by the boundary conditions in the form\cite{Ivanov}%
\begin{equation}
\xi _{S}(T)\frac{d}{dx}\Delta _{k}(z)=b(z)\Delta _{k}(z),~b(z)=\frac{\Sigma
_{1}(z)}{\Sigma _{2}(z)},  \label{BCGL}
\end{equation}%
\begin{equation}
\Sigma _{1}(z)=\sum_{\omega =-\infty }^{\infty }\xi _{S}(T)\frac{d}{dx}\frac{%
\Phi _{k}(z)}{\Omega ^{2}},~\Sigma _{2}(z)=\sum_{\omega =-\infty }^{\infty }%
\frac{\Phi _{k}(z)}{\Omega ^{2}},  \label{SigmaGL}
\end{equation}%
where $z=-d_{s},~0,~d_{F}.$ In typical experimental situation $\gamma
_{BI}\gg 1,$ $\gamma \sqrt{H}\gg 1$ and $d_{F}\sqrt{H}\gtrsim \xi _{F}.$ In
this case in the first approximation
\begin{equation*}
\Phi _{S}(d_{F})=0,~\Phi _{s}(0)=0,\frac{d}{dx}\Phi _{s}(-d_{s})=0
\end{equation*}%
and in the vicinity of interfaces
\begin{equation}
\Phi _{S}(x)=\Delta _{S}(x)=B_{S}\frac{(x-d_{F})}{\xi _{S}(T)}%
,~d_{F}\lesssim x\ll \xi _{S}(T),  \label{FiSas}
\end{equation}%
\begin{equation}
\Phi _{s}(x)=\Delta _{s}(x)=-B_{s}\frac{x}{\xi _{s}(T)},~-\xi _{S}(T)\ll
x\lesssim 0,  \label{FisAs1}
\end{equation}%
\begin{equation}
\Phi _{s}(x)=\Delta _{s}(x)=\Delta _{s}(-d_{s}),~-d_{s}\lesssim x\ll
-d_{s}+\xi _{S}(T),  \label{Fisas2}
\end{equation}%
where $B_{S},$ $B_{s},$ and $\Delta _{s}(-d_{s})$ are independent on $x$
constants. Substitution of the solutions (\ref{FiSas}) - (\ref{Fisas2}) into
(\ref{BCIFmod}), (\ref{Rel1}), (\ref{Rel2}) gives%
\begin{equation}
\Gamma _{BI}\xi _{S}(T)\frac{d}{dx}\Phi _{s}(-d_{s})=\Delta
_{s}(-d_{s})-\Delta _{0},  \label{next1}
\end{equation}%
\begin{equation}
\Phi _{S}(d_{F})=\frac{B_{s}}{\Gamma \sqrt{\Theta }\sinh \frac{d_{F}\sqrt{%
\widetilde{\Omega }}}{\xi _{F}}}+\frac{B_{S}\cosh \frac{d_{F}\sqrt{\Theta }}{%
\xi _{F}}}{\Gamma \sqrt{\Theta }\sinh \frac{d_{F}\sqrt{\Theta }}{\xi _{F}}},~
\label{next2}
\end{equation}%
\begin{equation}
\Phi _{s}(0)=\frac{B_{s}\cosh \frac{d_{F}\sqrt{\Theta }}{\xi _{F}}}{\Gamma
\sqrt{\widetilde{\Omega }}\sinh \frac{d_{F}\sqrt{\Theta }}{\xi _{F}}}+\frac{%
B_{S}}{\Gamma \sqrt{\Theta }\sinh \frac{d_{F}\sqrt{\Theta }}{\xi _{F}}},
\label{next3}
\end{equation}%
\begin{equation}
\Gamma _{BI}=\frac{\gamma _{BI}\xi _{S}}{\xi _{s}(T)},~\Gamma =\frac{\gamma
_{BI}\xi _{s}(T)}{\xi _{S}}.  \label{GammasB}
\end{equation}%
From definition (\ref{BCGL}), (\ref{SigmaGL}) of coefficients $~b(z)$ and
expressions (\ref{next1}) - (\ref{next3}) it follows that
\begin{equation}
\Gamma _{BI}\xi _{s}(T)\frac{d}{dx}\Delta _{s}(-d_{s})=-\left( \Delta
_{0}-\Delta _{s}(-d_{s})\right) ,  \label{BCGLF1}
\end{equation}%
\begin{equation}
\xi _{s}(T)\frac{d}{dx}\Delta _{s}(0)=-\frac{q+p}{2}\Gamma \Delta _{s}(0)-%
\frac{q-p}{2}\Gamma \Delta _{S}(d_{F}),  \label{BCGLF2}
\end{equation}%
\begin{equation}
\xi _{S}(T)\frac{d}{dx}\Delta _{S}(d_{F})=\frac{q+p}{2}\Gamma \Delta
_{S}(d_{F})+\frac{q-p}{2}\Gamma \Delta _{s}(0),  \label{BCGLF3}
\end{equation}%
where%
\begin{eqnarray}
p^{-1} &=&\frac{8}{\pi ^{2}}\re\sum_{\omega =0}^{\infty }\frac{1}{\Omega ^{2}%
\sqrt{\widetilde{\Omega }}\coth \frac{d_{F}\sqrt{\widetilde{\Omega }}}{2\xi
_{F}}},  \label{Sums} \\
q^{-1} &=&\frac{8}{\pi ^{2}}\re\sum_{\omega =0}^{\infty }\frac{1}{\Omega ^{2}%
\sqrt{\widetilde{\Omega }}\tanh \frac{d_{F}\sqrt{\widetilde{\Omega }}}{2\xi
_{F}}}.  \label{Sums1}
\end{eqnarray}

In considered limit both suppression parameters $\Gamma _{BI}\gg 1$ and $%
\Gamma \gg 1$ are large and from relations (\ref{BCIFmod}), (\ref{Rel1}), (%
\ref{Rel2}) in the first approximation on these parameters we get that the
boundary conditions (\ref{BCGLF1}) - (\ref{BCGLF3}) can be simplified to
\begin{equation}
\xi _{S}(T)\frac{d}{dx}\Delta _{s}(-d_{s})=0,~\Delta _{s}(0)=0,~\Delta
_{S}(d_{F})=0.  \label{BCGLs}
\end{equation}%
Taking into account that in this approximation supercurrent $j=0$ and $%
\Delta _{S}(\infty )=\Delta _{0}$ from (\ref{GLeq}), (\ref{BCGLs}) it
follows that
\begin{equation}
\Delta _{S}(x)=\delta _{S}(x)\exp \left\{ i\varphi \right\} ,~\delta
_{S}(x)=\Delta _{0}\tanh \frac{x-d_{F}}{\sqrt{2}\xi _{S}(T)},  \label{SolGL1}
\end{equation}%
while
\begin{equation}
\Delta _{s}(x)=\delta _{s}(x)\exp \left\{ i\chi \right\} ,  \label{SolGL2}
\end{equation}%
where $\delta _{s}(x)$ is the solution of transcendental equation
\begin{equation}
F\left( \frac{\delta _{s}(x)}{\delta _{s}(-d_{s})},\frac{\delta _{s}(-d_{s})%
}{\Delta _{0}\eta }\right) =-\frac{x\eta }{\sqrt{2}\xi _{s}(T)},~\eta =\sqrt{%
2-\frac{\delta _{s}^{2}(-d_{s})}{\Delta _{0}^{2}}}  \label{EqDsmall}
\end{equation}%
and $\delta _{s}(-d_{s})$ is a solution of the same equation at the SIs
boundary $x=-d_{s}$
\begin{equation}
K\left( \frac{\delta _{s}(-d_{s})}{\Delta _{0}\eta }\right) =\frac{d_{s}\eta
}{\sqrt{2}\xi _{s}(T)}.  \label{EqDsmallB}
\end{equation}%
Here $F(y,z)$ and $K(z)$ are the incomplete and complete elliptic integral
of the first kind respectively.

Substitution of (\ref{SolGL1}), (\ref{SolGL2}) into (\ref{BCGLF1}) - (\ref%
{BCGLF3}) gives that in the next approximation on $\Gamma _{BI}^{-1}$ and $%
\Gamma ^{-1}$%
\begin{equation}
J(-d_{s})=J_{G}\frac{\delta _{s}(-d_{s})}{\Gamma _{BI}\Delta _{0}}\sin
\left( \chi \right)  \label{JatIS}
\end{equation}%
\begin{equation}
J(0)=J(d_{F})=J_{G}\frac{\Gamma (p-q)}{2\Delta _{0}^{2}}\delta _{s}(0)\delta
_{S}(d_{F})\sin \left( \varphi -\chi \right) ,  \label{JatSF}
\end{equation}%
where
\begin{equation}
\delta _{s}(0)=-\frac{2b\left( q-p\right) \cos \left( \varphi -\chi \right)
+2a\left( q+p\right) }{\Gamma \left[ \left( q+p\right) ^{2}-\left(
q-p\right) ^{2}\cos ^{2}\left( \varphi -\chi \right) \right] },
\label{delta(0)}
\end{equation}%
\begin{equation}
\delta _{S}(d_{F})=\frac{2b\left( q+p\right) +2a\left( q-p\right) \cos
\left( \varphi -\chi \right) }{\Gamma \left( \left( q+p\right) ^{2}-\left(
q-p\right) ^{2}\cos ^{2}\left( \varphi -\chi \right) \right) },
\label{delta(df)}
\end{equation}%
are magnitudes of the order parameters at the FS interfaces and
\begin{equation}
a=-\delta _{s}(-d_{s})\sqrt{1-\frac{\delta _{s}^{2}(-d_{s})}{2\Delta _{0}^{2}%
}},~b=\frac{\Delta _{0}}{\sqrt{2}}  \label{derivaties}
\end{equation}%
Phase, $\chi ,$ of the order parameters of the s layer is determined from
equality of currents (\ref{JatIS}), (\ref{JatSF}).

\end{document}